\documentclass[twocolumn,tighten,twocolappendix]{aastex63}

\usepackage{color}

\shorttitle{Magnetic Field in a Extrasolar Gas Giant}
\shortauthors{Hori}
\begin{document}

\title{The Linkage between the Core Mass and the Magnetic Field of an Extrasolar Giant Planet from Future Radio Observations} %\footnote{Released on XXXX, XX, 2019}}

\correspondingauthor{Yasunori Hori}
\email{yasunori.hori@nao.ac.jp}

\author[0000-0003-4676-0251]{Yasunori Hori}
\affiliation{Astrobiology Center, 2-21-1 Osawa, Mitaka, Tokyo 1818588, Japan}
\affiliation{National Astronomical Observatory of Japan, 2-21-1 Osawa, Mitaka, Tokyo 1818588, Japan}



\begin{abstract}
Close-in gas giants are expected to have a strong magnetic field of $\sim 10-100$\,G. 
Magnetic fields in extrasolar giant planets are detectable by future radio observations in $\gtrsim 10$\,MHz and the spectropolarimetry of atomic lines. In contrast, the elusive interiors of exoplanets remain largely unknown.
%except for the existence of cloud/haze and chemical compositions in their upper atmospheres.
Here we consider the possibility of inferring the existence of the innermost cores of extrasolar giant planets through the detection of planetary magnetic fields. We simulated the long-term thermal evolution of close-in giant planets with masses of $0.2-10\,M_\mathrm{Jup}$ to estimate their magnetic field strengths. A young, massive gas giant tends to have a strong magnetic field. The magnetic field strength of a hot Jupiter is insensitive to its core mass, whereas the core strongly affects the emergence of a planetary dynamo in a hot Saturn. No dynamo-driven magnetic field is generated in a hot Saturn with no core or a small one until $\sim 10-100$\,Myr if metallization of hydrogen occurs at $\gtrsim 1-1.5$\,Mbar.
%A highly-luminous, massive gas giant maintains a strong magnetic field until it achieves the thermal equilibrium under stellar irradiation.
The magnetic field strength of an evolved gas giant after $\sim 100\,\mathrm{Myr}$ is almost independent of the stellar incident flux.
Detecting the magnetic field of a young, hot Saturn as a good indicator of its core may be challenging because of the weakness of radio signals and the shielding effect of plasma in the Earth's ionosphere. Hot Jupiters with $\gtrsim 0.4\,M_\mathrm{Jup}$ can be promising candidates for future ground-based radio observations.
\end{abstract}

\keywords{Exoplanets; Planetary Interior; Magnetic fields}


\section{Introduction} \label{sec:intro}

Planetary magnetic fields play an important role in shielding planetary atmospheres from the invasion of high-energy particles via a stellar wind and coronal mass ejections and then inhibiting the nonthermal mass loss from a planet.
Magnetic fields of planets are generated by convective motions in an electrically conducting fluid, the so-called planetary dynamos.
The resulting magnetic fields are produced by the electric currents in iron-nickel fluid in the Earth's outer core, liquid metallic hydrogen in gas giants, and ionic water in ice giants. 
In the solar system, Mercury, Earth, the four giant planets, and maybe Ganymede have dynamo-driven fields. For example, Jupiter has the most intense magnetic field at $\sim11.8$ times that of Earth at the south pole \citep{2017Sci...356..821B}. 

A theoretical relationship between planetary magnetic fields on the surface and the physical properties of planets predicted that close-in extrasolar gas giants as well as Jupiter have $\sim 10$--100\,G \citep{2010A&A...522A..13R,2017ApJ...849L..12Y}. Stars hosting hot Jupiters show flux changes in the cores of Ca\,II H and K lines on a timescale comparable to the planet's orbital period \citep[e.g.][]{2003ApJ...597.1092S,2005ApJ...622.1075S}. Such modulations of chromospheric emissions are interpreted as evidence of magnetic star-planet interactions (SPI). Recently, \citet{2019NatAs...3.1128C} reported that flux variations in the Ca II K line (3933.66\,$\mathrm{\AA}$) were synchronized to the planet's orbital phase for four hot Jupiter systems: HD\,179949, HD\,189733, $\tau$\,Boo, and $\upsilon$\,And. Assuming that chromospheric flux changes are involved in energy dissipation in a magnetic loop interconnecting the base of the stellar corona with the planetary surface \citep{2013A&A...557A..31L}, they found that the field strengths of the four hot Jupiters range from $\sim20$ to 100\,G. The estimated magnetic fields of hot Jupiters are consistent with the scaling law relating the magnetic field strength to the heat flux in gas giants \citep{2009Natur.457..167C}.

Measuring the properties of planetary magnetic fields provides valuable insights into the interiors of planets. Magnetic fields split the atomic energy levels in multi-substates due to the Zeeman effect or Paschen-Back effect, which causes polarization in spectral lines. In the presence of a weak magnetic field, atomic-level linear polarization is modified through processes known as the Hanle effect. \citet{2020ApJ...890...88O} proposed a spectropolarimetric method for detecting magnetic fields in exoplanets, using linear or circular polarization in the triplet line (1083\,nm) of He I escaping from close-in exoplanets within 0.1\,au during the transits \citep[see also][]{2018ApJ...855L..11O,2019ApJ...881..133O}.

Another diagnostic aid for magnetic fields in planets utilizes the radio emission. Nonthermal radio emissions from planets occur by interactions between planetary magnetic fields and high-energy, charged particles coming from a stellar wind and coronal mass ejections \citep[e.g.][]{2005A&A...437..717G,2007P&SS...55..618G}. The synchrotron radiation is emitted from accelerated electrons in a magnetic field. The cyclotron radiation from electrons moving along the magnetic field of a planet, that is, the auroral radio emission, is attributed to the electron-cyclotron maser instability. The latter radiation is strongly beamed in the forward direction within a cone, and its frequency is given by $f_{\rm cyc} = eB/(2\pi m_{\rm e}c) \sim 2.8(B/1\,{\rm G})$\,MHz, where $e$ is the elementary charge, $m_{\rm e}$ is the electron mass, $c$ is the speed of light, and $B$ is the magnetic field strength of a planet. The expected radio fluxes from hot Jupiters range from $\sim \mu$\,Jy to m\,Jy in 10\,MHz to a few 100\,MHz \citep[e.g.][]{2007A&A...475..359G}.\footnote{
Radio emissions in the frequency range of $\lesssim 8.98$\,MHz are shielded by plasma in the Earth's ionosphere.
}
No direct detection of magnetic fields in exoplanets has been reported. Next-generation radio telescope projects such as the Square Kilometre Array (e.g. SKA-low) will enable us to capture radio signals from nearby hot Jupiters.

Planetary magnetic fields are related to the thermal state and internal compositions of planets.
The bulk composition and formation processes of giant planets are closely linked. Some close-in giant planets such as HD\,149026b have an extremely large mean density, which suggests the enrichment in heavy elements. Determining the interior structure of close-in giant planets, however, requires high-precision measurements of their gravity field. 
In contrast, magnetic fields in exoplanets are expected to be detected by future observations of auroral radio emissions in $\gtrsim 10$\,MHz and the spectropolarimetry of atomic lines such as the He I triplet. In this paper, we aim to infer the existence of the innermost cores of extrasolar giant planets from the strength of planetary magnetic fields.
In Section 2, we describe an empirical relationship between magnetic fields in planets and their thermodynamic properties and our numerical recipes for calculating the long-term thermal evolution of a planet. We present the strength of magnetic fields in evolving giant planets in Section 3. We discuss constraints on the core masses of nearby close-in gas giants from radio observations in Section 4. We summarize this study in the last section. 

\section{Method} \label{sec:method}

\subsection{Planetary Magnetic Field}

%A planetary dynamo requires for the Elsasser number $\Lambda = \sigma B^2/(2\rho \Omega)$ (i.e., the ratio of Lorentz and Coriolis forces) of the order unity, which is often called the Elsasser number rule \citep[e.g.][]{1984AN....305..257S}, where $\sigma$ is the electrical conductivity, $B$ is the magnetic field strength, $\rho$ is the fluid density in the dynamo, and $\Omega$ is the rotation rate. For example, we have $\Lambda \sim$ 0.1 for Earth and 1 for Jupiter \citep[e.g. see][]{2011PEPI..187...92S}.
Many scaling rules for the field strength in a planetary dynamo have been proposed using the rotation rate and the heat flux in the dynamo region \citep[e.g.][]{1986JGR....9111003C,2002Icar..157..426S}.
%, e.g., the well-known mixing length theory. 
Recently, numerical dynamo simulations of rapidly rotating bodies suggested that the magnetic field strength is 
independent of the rotation rate and can be controlled by the buoyancy flux \citep{2006GeoJI.166...97C,2013Icar..225..185Y,2013ApJ...774....6Y}.
\citet{2009Natur.457..167C} showed that the magnetic field strength in a planet is empirically estimated from a scaling law based on the energy flux available for a planetary dynamo.\footnote{
The heat flux can be converted to magnetic energy to maintain the dynamo against ohmic dissipation. This concept, specifically a linear dependence of $B^2$ on the thermal flux, was considered in modelling the magnetism of the thermally evolving terrestrial planets \citep{1983Icar...54..466S}.}
A theoretical consideration assuming a MAC balance also proposed a similar powered-based scaling law \citep{2013GeoJI.195...67D}.

We consider that the heat flux by convection compensates for energy loss by ohmic dissipation in the dynamo region. 
The energy-flux-based magnetic energy density is given by
\begin{equation}
    \frac{B^2}{2\mu_0} \propto f_{\rm ohm} \rho^{1/3} 
                                \left( \frac{q_{\rm c} L}{H_{T}} \right)^{2/3},
    \label{eq:B}
\end{equation}
where $\mu_0$ is the permeability of free space, $q_{\rm c}$ is the convective heat flux, $f_{\rm ohm}$ is the fraction of the thermal flux converted to magnetic energy, and $H_T$ is the temperature scale height \citep[see][for details]{2010SSRv..152..565C}.
According to \citet{2009Natur.457..167C}, we obtain the mean squared magnetic field $\langle B \rangle$ by radially averaging equation (\ref{eq:B}) over the volume $V$ of the spherical shell:
\begin{eqnarray}
    \frac{\langle B \rangle^2}{2\mu_0} &=& c f_{\rm ohm} \langle \rho \rangle^{1/3}  (F q_0)^{2/3}, \label{eq:ave-B} \\
    F^{2/3} &=& \frac{1}{V} \int^{R}_{r_{\rm i}} \left(\frac{q_{\rm c}(r)}{q_0}\frac{L(r)}{H_T(r)} \right)^{2/3} \nonumber
                \left(\frac{\rho(r)}{\langle \rho \rangle} \right)^{1/3} 4 \pi r^2 dr,
\end{eqnarray}
where $c$ is a constant, $F$ represents the averaging factor for radially varying properties (which is of the order unity for Earth and Jupiter),
$r$ is the radial distance from the center of a planet,
$R$ and $r_{\rm i}$ are the outer radius and the inner radius of the convective layer, $\langle \rho \rangle$ is the mean density of a fluid in the convective layer, $L$ is the length scale of the largest convective structure,
$q_0$ defines the reference flux as $q_0 = q_{\rm i,c} (r_{\rm i}/R)^2$, and $q_{\rm i,c}$ denotes the effective flux on the inner boundary of the convective layer.
The scaling law with $c = 0.63$ taken from dipolar dynamo simulations and $f_{\rm ohm} = 1$ is in agreement with the observed magnetic fields of Earth and Jupiter and the mean internal field strengths of the classical T Tauri stars, old M dwarfs, and rapidly rotating solar-type stars with rotational periods of $\lesssim 4$\,days \citep{2009Natur.457..167C}.
\footnote{
\citet{2009ApJ...697..373R} also derived a simple scaling law of magnetic field strengths for young, low-mass stars from their dynamo simulations:
\begin{equation}
    Bf \sim 4.8^{+3.2}_{-2.8} \left(\frac{ML^2}{R^7}\right)^{1/6}\,{\rm kG},
    \label{eq:emp-B}
\end{equation}
where $B$ is the mean magnetic field strength at the dynamo surface, and $M$, $L$, and $R$ are the mass, luminosity, and radius in units of solar values, respectively. The filling factor $f$, which is the fraction of the surface area filled with the magnetic field, is assumed to be unity. 
The strength of the dipole field on the planetary surface at the pole $B^{\rm pole}_{\rm dipole}$ for a giant planet is related to $B$ \citep{2010A&A...522A..13R} by
\begin{equation}
    B^{\rm pole}_{\rm dipole} = \frac{B}{\sqrt{2}} \left(1-\frac{0.17}{M_{\rm p}/M_{\rm Jup}} \right)^3,
\end{equation}
where $M_{\rm p}$ is the planetary mass.}

The dynamo surface is defined as the top of a metallic hydrogen core, namely, the emergence of metallic hydrogen in the convective layer. Both high-pressure experiments and ab initio calculations for hydrogen-helium mixtures predict that the (continuous?) transition from molecular to metallic hydrogen occurs at $T \gtrsim 1,500-2,000\,{\rm K}$ and $P \gtrsim 1-2\,{\rm Mbar}$ \citep[e.g.][]{2012PhRvB..86n4115L,2013PhRvL.110f5702M}. In this paper, we consider a critical point for the phase transition to be located near $T = 2,000{\rm K}$ and $P = 2{\rm Mbar}$. Metallic hydrogen, however, may appear under $\lesssim 2$\,Mbar \citep[e.g.][]{2012ApJS..202....5F}. We also consider a lower critical pressure at which metallization of hydrogen occurs in Section 3.1.

\subsection{Thermal Evolution of a Planet}

An initially hot, inflated giant planet continues to cool and gravitationally contract over time. We simulate the long-term thermal evolution of an irradiated giant planet for 10\,Gyr using the stellar evolution code {\tt MESA} \citep{2011ApJS..192....3P}.
The general-purpose {\tt MESA} code was originally developed for one-dimensional stellar evolution calculations. The extended {\tt MESA} code \citep{2013ApJS..208....4P} is capable of modeling the evolution of giant planets with masses larger than 0.1 times the mass of Jupiter.
The applicability of {\tt MESA} to irradiated giant planets has been tested by other codes for calculating the evolution and interiors of giant planets, such as {\tt CEPAM} \citep{1995A&AS..109..109G}.
%has been adopted in previous works for evolutionary calculations of irradiated planets \citep[e.g.][]{2013ApJ...775..105O}. 
In this study, a planet receives an incident flux of 10--1000\,$F_\oplus$ from its parent star at the substellar point, where $F_\oplus$ is the incident flux received by Earth.
The heating effect of a planetary surface by stellar irradiation in {\tt MESA} adopts the $F_\star$--$\Sigma_\star$ method, which gives good agreement with an analytical semigrey model for plane-parallel, irradiated planetary atmospheres \citep{2010A&A...520A..27G}. Since the thermal evolution of an irradiated planet with an equilibrium temperature of $\lesssim$ 1300\,K is not strongly influenced by the choice of $\Sigma_\star$ value \citep{2015ApJ...813..101V}, we use $\Sigma_\star = 330$\,g\,cm$^{-2}$, corresponding to $\kappa_\nu (= 2/\Sigma_{\star}) = 6 \times 10^{-3}$\,cm$^{2}$\,g$^{-1}$ in \citet{2010A&A...520A..27G}.
The magnetic field strength of a thermally evolving gas giant is calculated using equation (\ref{eq:ave-B}).
The thermodynamic properties (e.g. the convective heat flux $q(r)$) required for equation (\ref{eq:ave-B}) are estimated from {\tt MESA} calculations (see the Appendix A for more details).

We consider close-in gas giants with masses of 0.2--10\,$M_{\rm Jup}$ where $M_{\rm Jup}$ is Jupiter's mass. Each planet is assumed to have a core-envelope structure. The mass of an innermost core ranges from 0 to 10\,$M_\oplus$ ($M_\oplus$ is the Earth's mass).
The core is assumed to have a constant density of 5\,g\,cm$^{-3}$. 
The H$_2$/He envelope of a planet has a uniform solar composition
(see Section 3.3 for the effect of the metallicity ([Fe/H]) in the envelope on the strength of a planetary magnetic field).
{\tt MESA} employs SCvH EoS for hydrogen and helium \citep{1995ApJS...99..713S} and the opacity tables of \citet{2005ApJ...623..585F} and \citet{2008ApJS..174..504F}. A gas giant is assumed to have a relatively large radius of 2\,$R_{\rm Jup}$ ($R_{\rm Jup}$ is Jupiter's radius) shortly after the formation, which mitigates {\tt MESA} convergence problems \citep{2013ApJS..208....4P}. An initially hot gas giant rapidly radiates heat into space, and its envelope sharply shrinks on a $1-10$\,Myr timescale. As a result, the thermal evolution of the irradiated gas giant after $\sim 10$\,Myr is almost independent of its initial radius.

\begin{figure}[t]
    \centering
    \includegraphics[clip,scale=0.32]{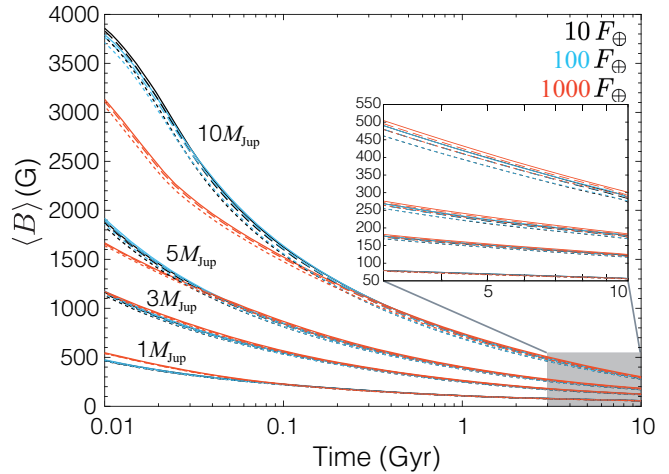}
    \caption{Magnetic field strength ($\langle B \rangle$) of close-in gas giants with four different masses (1, 3, 5, and 10\,$M_{\rm Jup}$). We consider three cases in which a planet has no core (solid line), a $3M_\oplus$ core (dashed), and a $10M_\oplus$ core (dotted). Each color corresponds to the stellar flux that a planet receives: 10 (black), 100 (cyan), and 1000\,$F_\oplus$(red).}
    \label{fig:Bdy-massive}
\end{figure}

\begin{figure}[t]
    \centering
    \includegraphics[clip,scale=0.32]{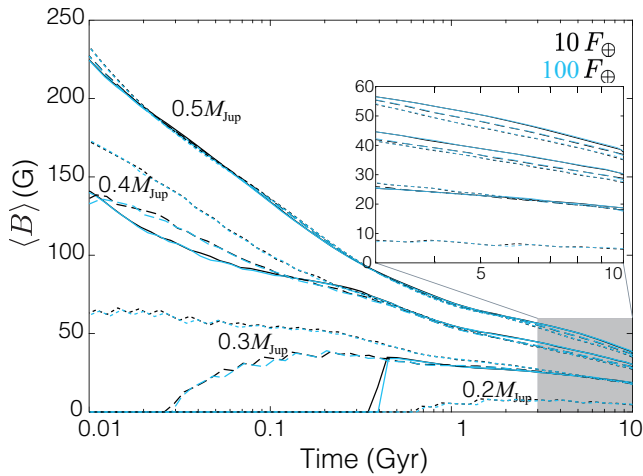}
    \caption{Same as Figure \ref{fig:Bdy-massive} but for 0.2, 0.3, 0.4, and  0.5\,$M_{\rm Jup}$.}
    \label{fig:Bdy-low}
\end{figure}

\section{Results}\label{sec:result}

The physical and thermodynamic properties of a fluid in the dynamo region in a planet are derived from evolutionary calculations over time.
Then we estimate the strength of a planetary magnetic field using equation (\ref{eq:ave-B}).

\subsection{Planet mass}

Figure \ref{fig:Bdy-massive} shows that the magnetic field strength of close-in gas giants with $\gtrsim 1\,M_{\mathrm{Jup}}$ (hereafter referred to as a "Jupiter-mass planet") decreases with time. The magnetic field strengths of irradiated gas giants, which have fully convective interiors, are determined by the mean density of a fluid and heat flux in the dynamo region (see equation (\ref{eq:ave-B})). 
A larger heat flux in the dynamo region of a more massive planet yields a stronger magnetic field.
As an initially hot, Jupiter-mass planet contracts, the location of the dynamo surface moves downward slowly. The dynamo region in the envelope of a massive gas giant shrinks over time, whereas the mass of a dense fluid in the dynamo region slightly increases. A planet continues to radiate heat away into space, which reduces the energy flux to maintain the dynamo against ohmic dissipation. Since the rate of energy loss by radiation overwhelms the compression effect of fluid in the dynamo region, the magnetic field strength of a Jupiter-mass planet drops monotonically. A Jupiter-mass planet with a larger core also tends to have a slightly weaker magnetic field because of a shorter cooling timescale.

\begin{figure}[t]
    \centering
    \includegraphics[clip,scale=0.32]{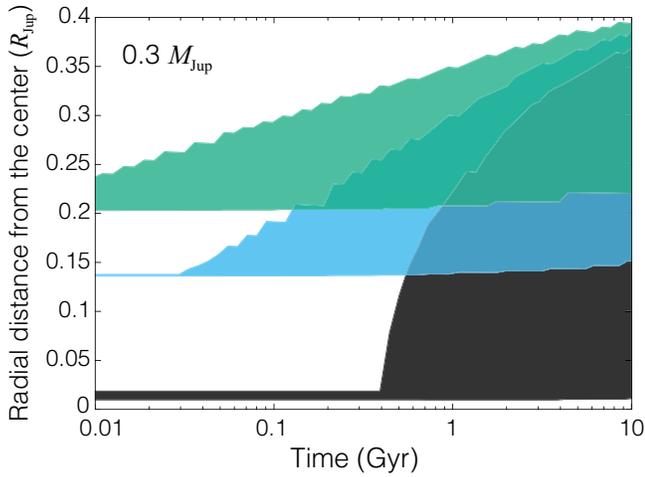}
    \caption{Dynamo region in a 0.3\,$M_{\rm Jup}$ planet with no core (black), a $3M_\oplus$ core (cyan), and a $10M_\oplus$ core (green). The incident flux on the planet is 10\,$F_\oplus$. The bottom of each dynamo region is located at the core surface.}
    \label{fig:dynamo}
\end{figure}

The core mass strongly affects the emergence of the dynamo region in a close-in gas giant with $\lesssim 0.3\,M_{\mathrm{Jup}}$ (hereafter called a "Saturn-mass planet"). The interior of a Saturn-mass planet is hot enough ($> 2000\,{\mathrm K}$) in the early stage ($\sim 1-10\,\mathrm{Myr}$) of its thermal evolution. The metallization of hydrogen does not occur because the pressure is lower than $\sim 2\,\mathrm{M\,bar}$ even near the core-envelope boundary. In such a Saturn-mass planet, the magnetic field is never generated by the dynamo mechanism until its interior satisfies the conditions of pressure and temperature for hydrogen metallization. Figure \ref{fig:Bdy-low} shows the magnetic field strengths of four Saturn-mass planets that have no core, a $3\,M_\oplus$ core, and a $10\,M_\oplus$ core as a function of time.
The magnetic field strengths of close-in gas giants with $\gtrsim 0.4\,M_{\mathrm{Jup}}$ monotonically decrease with time, as well as Jupiter-mass planets, whereas 0.2\,$M_{\mathrm {Jup}}$ and 0.3\,$M_{\mathrm {Jup}}$ planets without a massive core have no dynamo-driven magnetic field until 30--300\,Myr. 
As an initially inflated Saturn-mass planet gravitationally contracts, the dynamo region first appears near the core surface and expands upward with time (see Figure \ref{fig:dynamo}). If a 0.3\,$M_{\mathrm {Jup}}$ planet has a $10\,M_\oplus$ core, the envelope contraction proceeds rapidly, and then it creates a dynamo region to generate a persistent magnetic field immediately after the formation. Thus, the magnetic field strength of a young Saturn-mass planet is closely related to its core mass. Also, the more massive the Saturn-mass planets are, the stronger their magnetic fields are.

%{\bf Magnetic field strengths on the surface ($\langle B_{\mathrm{s}}\rangle$) of four Jupiter-mass planets at the pole, where $\langle B_{\mathrm{s}}\rangle \propto \langle B \rangle (R_{\mathrm{dyn}}/R_{\mathrm{s}})^3$, $R_{\mathrm{dyn}}$ is the location of the dynamo surface, and $R_{\mathrm{s}}$ is the planet's radius}

\begin{figure}[t]
    \centering
    \includegraphics[clip,scale=0.32]{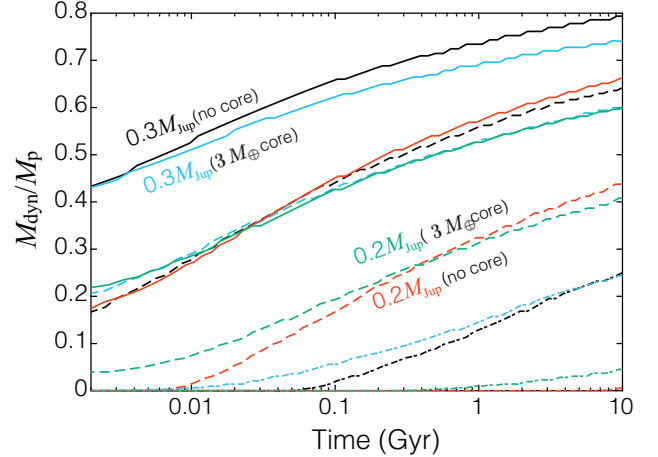}
    \caption{
    Mass enclosed in the dynamo region ($M_\mathrm{dyn}$) with respect to the total mass ($M_\mathrm{p}$) of a hot Saturn as a function of time: 0.3\,$M_{\rm Jup}$ planets with no core (black) and a $3M_\oplus$ core (cyan) and 0.2\,$M_{\rm Jup}$ planets with no core (red) and a $3M_\oplus$ core (green). Each planet receives $10\,F_\oplus$ from its central star. The phase transition of molecular to metallic hydrogen occurs at $0.25$\,Mbar and $4000$\,K (solid line), $1$\,Mbar and $2000$\,K (dashed), and $1.5$\,Mbar and $1500$\,K (dashed-dotted), respectively. 
    }
    \label{fig:t-Mdyn}
\end{figure}

\begin{figure}[t]
    \centering
    \includegraphics[clip,scale=0.32]{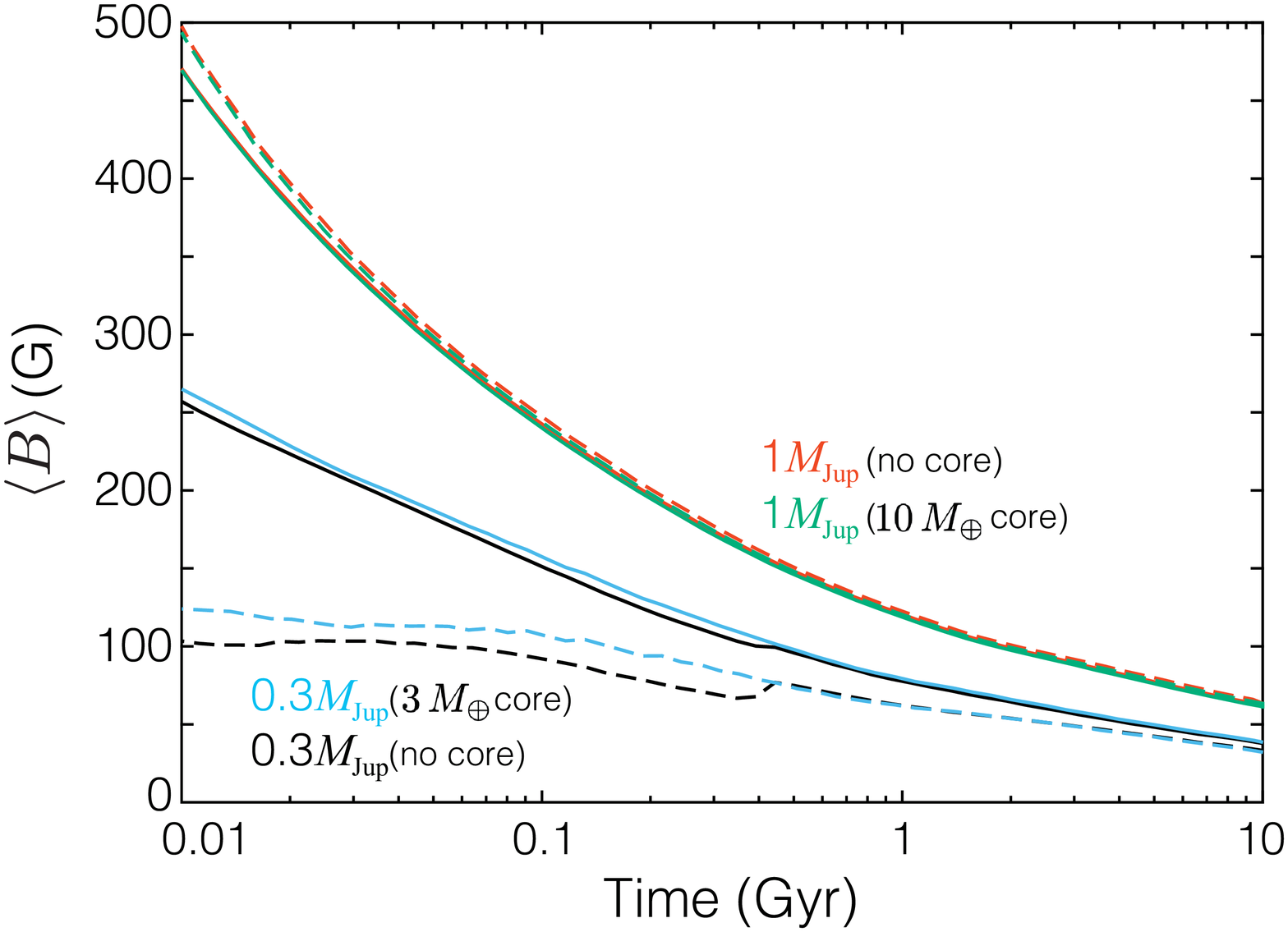}
    \caption{
    Magnetic field strengths in 0.3\,$M_{\rm Jup}$ planets with no core (black) and a $3 M_\oplus$ core (cyan) and 1\,$M_{\rm Jup}$ planets with no core (red) and a $10 M_\oplus$ core (green) in which metallization of hydrogen occurs at $0.25$\,Mbar and $4000$\,K (solid line) and $1$\,Mbar and $2000$\,K (dashed).
    Each planet has an incident flux of $10\,F_\oplus$ from its parent star.
    }
    \label{fig:B-Ptran}
\end{figure}

Metallization of hydrogen may occur under $\lesssim 2$\,Mbar. Molecular dynamics simulations of liquid hydrogen under high pressure and temperature \citep{2010PhRvL.104f5702T} suggested that the phase boundaries of molecular-to-metallic hydrogen are located near 0.25, 1, and 1.5\,Mbar for 4000\,K, 2000\,K, and 1500\,K, respectively. Figure \ref{fig:t-Mdyn} shows the mass fractions of the metallic hydrogen region (i.e., the dynamo region) in a $0.2\,M_\mathrm{Jup}$ planet and a $0.3\,M_\mathrm{Jup}$ planet for 10Gyr. A critical pressure at which hydrogen metallization occurs determines the emergence of a dynamo region in a hot Saturn because its interior is always hotter than $4000$\,K at $0.25-1.5$\,Mbar. 
Although metallization of hydrogen at a lower pressure enlarges the electrically conducting region required for a dynamo, a hot Saturn with a larger core yields a less massive dynamo region because the envelope of a hot Saturn with a core becomes compressed.
The appearance of metallic hydrogen at $\lesssim 1-1.5$\,Mbar allows a hot Saturn with no core or a smaller core to generate a dynamo-driven magnetic field right after its formation. This can be supported by the relevant criterion for a dynamo: the magnetic Reynolds number $R_\mathrm{m} = U L \mu_\mathrm{o} \sigma$, where $U$ is the flow velocity, $L$ is the characteristic length scale of the flow, $\mu_\mathrm{o}$ is the permeability of vacuum, and $\sigma$ is the electrical conductivity. For a hot Saturn or Jupiter having a metallic hydrogen envelope, $U \sim 0.1-1$\,m\,s$^{-2}$ and $L \sim 10^7$\,m give $R_\mathrm{m} \sim 10-100 \sigma$\,Sm$^{-1}$. If metallic hydrogen has $\sigma \gtrsim 10^3-10^4$\,Sm$^{-1}$ \citep{2012ApJS..202....5F}, the metallic hydrogen region in a hot Saturn can generate a self-sustained dynamo \citep[e.g.][]{2006GeoJI.166...97C}. Figure \ref{fig:B-Ptran} shows magnetic field strengths of a hot Saturn and a hot Jupiter in which the metallization of hydrogen occurs at $\lesssim 1$\,Mbar. A hot Saturn with no core or a small core having an envelope occupied by metallic hydrogen can yield a magnetic field stronger than $\sim 50$\,G. The magnetic field strength of a hot Saturn depends on the mass of an inner core, whereas that of a hot Jupiter is insensitive to the existence of a core.

\begin{figure}[t]
    \centering
    \includegraphics[clip,scale=0.32]{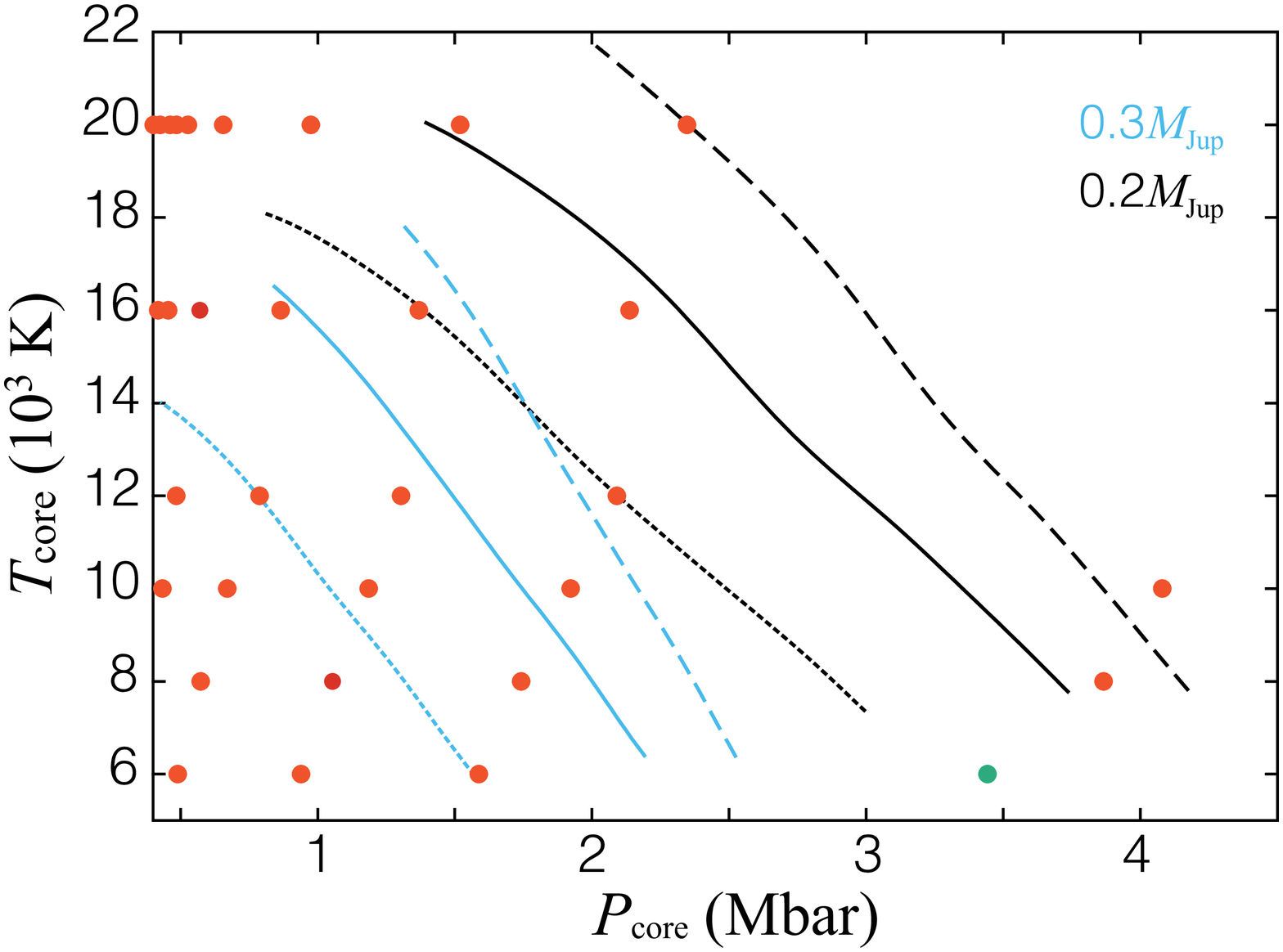}
    \caption{
    Pressure ($P_\mathrm{core}$) and temperature ($T_\mathrm{core}$) at the core surface in 0.2\,$M_{\rm Jup}$ planets (black) and 0.3\,$M_{\rm Jup}$ planets (cyan) with no core (dotted line), a $3 M_\oplus$ core (solid), and a $10 M_\oplus$ core (dashed) for 10\,Gyr, respectively. Filled circles correspond to phase diagrams from ab initio simulations of water under high pressure \citep{2009PhRvB..79e4107F}: plasma (red) and superionic water (green).
    All of the planets are assumed to receive $10\,F_\oplus$ from their parent star.
    }
    \label{fig:Pc-Tc}
\end{figure}

Known hot Jupiters ($\gtrsim 1\,M_{\mathrm{Jup}}$) orbiting stars with ages of $1-10\,{\mathrm{Gyr}}$ likely have stronger surface magnetic fields than the present-day Jupiter.
Actual mean field strengths on the surface of irradiated giant planets are likely to be weaker than the internal ones, that is, $\langle B \rangle$ in equation (\ref{eq:ave-B}).
The expected magnetic field strength for $0.5-5\,M_{\mathrm{Jup}}$ planets that receive $100-1000\,F_\oplus$ from their host stars aged $\gtrsim 1\,{\mathrm{Gyr}}$ range from 10 to 300\,G, which are consistent with those of four hot Jupiters in \citet{2019NatAs...3.1128C}.
%In contrast, the surface of close-in gas giants with $0.5\,M_{\mathrm{Jup}}$ shows a steady magnetic field for 10\,Gyr (see Figure \ref{fig:Bs-low}). As seen in Figure \ref{fig:Bdy-low}, magnetic field strengths at the dynamo surface of such planets monotonously decrease with time. However, as a $0.5\,M_{\mathrm{Jup}}$ planet cools and contracts, the dynamo region gradually expands. Increasing the ratio of $R_{\mathrm{dyn}}/R_{\mathrm{s}}$ counteracts the decrease in $ B_{\mathrm{dyn}}$ for $0.5\,M_{\mathrm{Jup}}$. 
Less massive gas giants, specifically, hot Saturns with masses of $\lesssim 0.3\,M_\mathrm{Jup}$ may be non-magnetized planets for 0.1--1\,Gyr unless they have a large core or metallization of hydrogen occurs at $\lesssim 1-1.5$\,Mbar. In other words, a young, hot Saturn with no core or a small core might have a very weak dynamo-driven magnetic field at the surface.
Note that icy material under high pressure and temperature plays a role in generating a dynamo-driven magnetic field, as seen in Uranus and Neptune. Plasma or ionic water can be an electrically conducting fluid. Figure \ref{fig:Pc-Tc} shows the pressure and temperature at the core surface in hot Saturns with no core, a $3 M_\oplus$ core, and a $10 M_\oplus$ core. Phase diagrams of water from ab initio simulations of \citet{2009PhRvB..79e4107F} are also shown in Figure \ref{fig:Pc-Tc}. If the inner core of a close-in planet is composed of icy material (i.e., water), water in an icy core exists as a plasma for 10\,Gyr. Thus, the existence of an icy core in a hot Saturn and hot Jupiter contributes to the strength of a dynamo-driven magnetic field.

\subsection{Stellar incident flux}

\begin{figure}
    \centering
    \includegraphics[clip,scale=0.32]{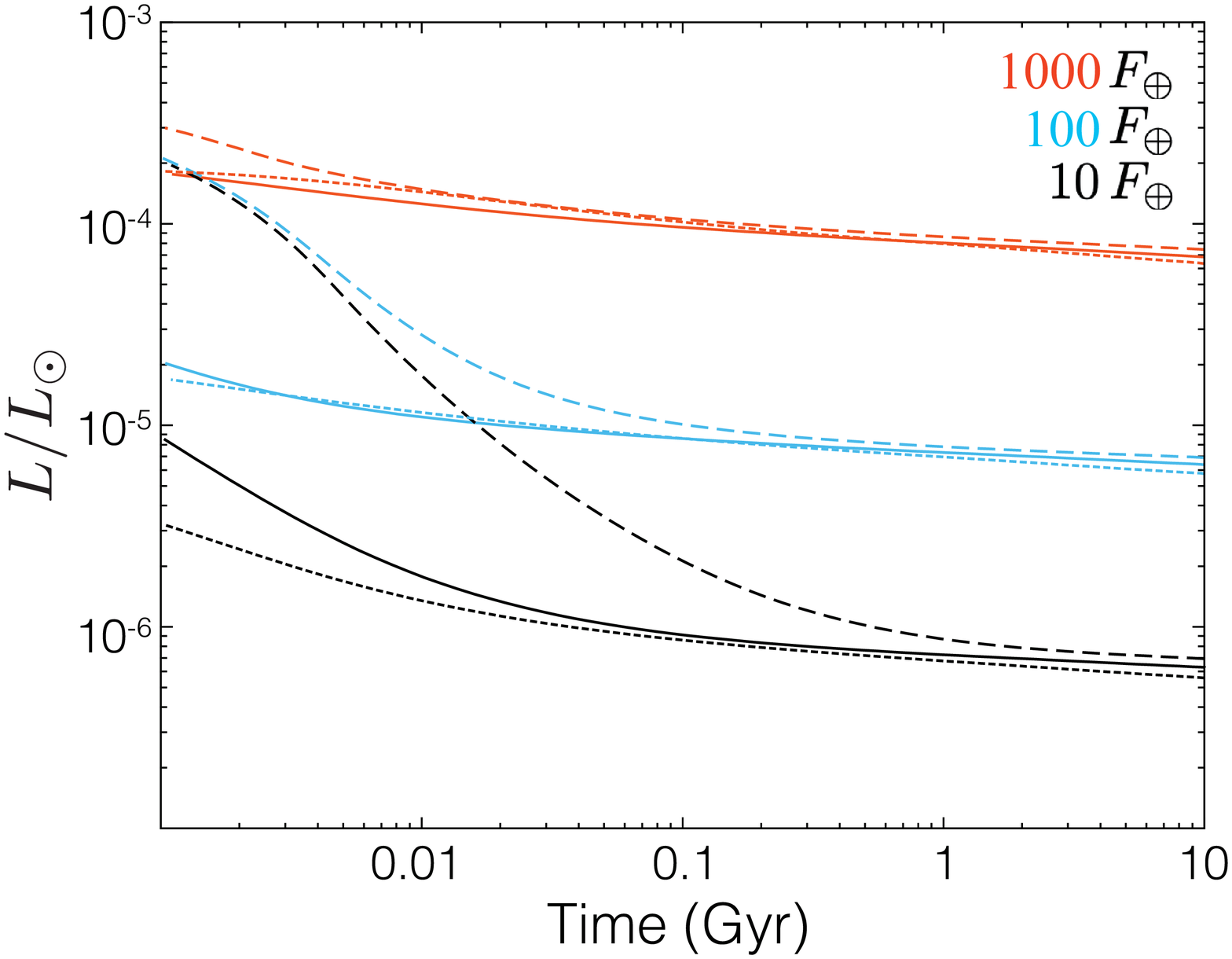}
    \caption{Intrinsic luminosity of 0.5 (dotted line), 1 (solid), and $5\,M_{\mathrm{Jup}}$ planets (dashed) with a $10\,M_\oplus$ core. Each color corresponds to the stellar flux that a planet receives: 10 (black), 100 (cyan), and 1000\,$F_\oplus$(red).}
    \label{fig:t-L}
\end{figure}

The hot state of a close-in gas giant continues until 10--100\,Myr. Figure \ref{fig:t-L} shows the intrinsic luminosity of 0.5, 1, and $5\,M_{\mathrm{Jup}}$ planets with a $10M_\oplus$ core. A massive planet maintains a highly luminous state for a long time. Such a self-luminous planet has to lose a large amount of internal heat to achieve thermal equilibrium under less intense stellar irradiation. As a result, a young, massive close-in gas giant with $\gtrsim 5\,M_{\mathrm{Jup}}$ that receives a low incident flux from its host star tends to have a strong magnetic field (see Figure \ref{fig:Bdy-massive}). As the atmosphere of an irradiated planet asymptotically approaches to the state of thermal equilibrium, the heat flow from the convective interior declines. The magnetic field strength of a planet in thermal equilibrium is eventually insensitive to the stellar incident flux (see Figures \ref{fig:Bdy-massive} and \ref{fig:Bdy-low}).

%\subsection{Surface magnetic field}
%Cauley+:stars aged 1Gyr, incident flux = 350-1,400 Fe, planetary mass = 0.7-4Mjup --- 10-100 G

\subsection{Metallicity}

\begin{figure}
    \centering
    \includegraphics[clip,scale=0.32]{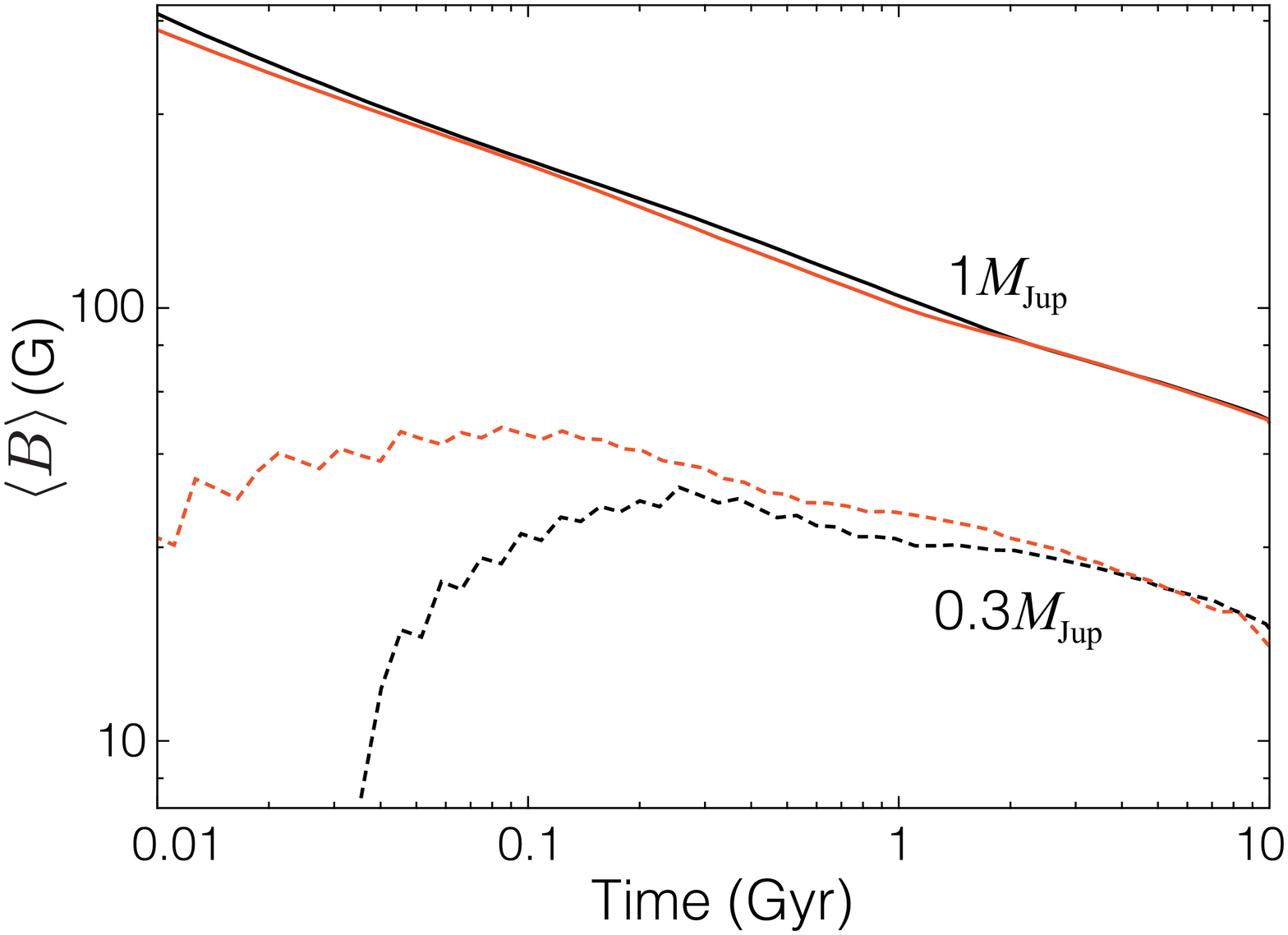}
    \caption{Magnetic field strength of $0.3\,M_\mathrm{Jup}$ with $M_z = 6\,M_\oplus$ (dashed) and $1\,M_\mathrm{Jup}$ with $M_z = 20\,M_\oplus$ (solid), where $M_z$ is the total amount of heavy elements. Two interior models of each planet are assumed: (red) both planets have a metal-free envelope on a $M_z$ core and (black) $M_z$ is divided evenly between the envelope and the core.}
    \label{fig:B-Z-Mp}
\end{figure}

The cooling of a planet is controlled by the distribution of heavy elements within itself.
Observational constraints on Jupiter and Saturn suggest that their interiors are enriched by heavy elements at $Z \sim 0.1\%-0.2\%$ \citep[e.g.][]{2013ApJ...767..113H,2017GeoRL..44.4649W}. Jupiter and Saturn likely have a core surrounded by a metal-rich envelope.
Figure \ref{fig:B-Z-Mp} demonstrates the distribution of heavy elements within a Saturn-mass planet and a Jupiter-mass planet on their magnetic field strengths. We considered two core-envelope models for a $0.3\,M_\mathrm{Jup}$ planet with $M_z = 6\,M_\oplus$ and a $1\,M_\mathrm{Jup}$ planet with $M_z = 20\,M_\oplus$ respectively, where $M_z$ is the total amount of heavy elements: (1) all heavy elements are located in the core and (2)  $M_z$ is divided evenly between the envelope and the core.
Internal heat can be mainly transferred upward by convection even in the metal-free envelopes of the two planets. The condition for a planetary dynamo is satisfied by the metallization of hydrogen. The core gravity has a non-negligible contribution to the $P-T$ profile in the envelope of a Saturn-mass planet. Since the dynamo region lies deep within a Saturn-mass planet, 
the mass of heavy elements in the central region affects the emergence of the magnetic field and the strength, as seen in Figure \ref{fig:Bdy-low}. On the other hand, the dynamo surface of a Jupiter-mass planet is located near the planetary surface, regardless of its core mass. As a result, the magnetic field strength of a Jupiter-mass planet is insensitive to the distribution of heavy elements within itself (see also Figure \ref{fig:Bdy-massive}).

The metallicity of stars harboring close-in gas giants ranges from -0.3 to +0.4. Giant planets around metal-rich stars appear to have large heavy-element masses \citep{2006A&A...453L..21G,2011ApJ...736L..29M}. Also, there is a negative correlation between the heavy-element enrichment in the planet relative to its parent star and the planetary mass \citep{2016ApJ...831...64T}. Upper limits on atmospheric metallicities for transiting giant planets are estimated to be $Z_\mathrm{atm}/Z_\odot \gtrsim 10$ \citep{2019ApJ...874L..31T}. 
%, which are consisten with 透過分光でもmetal-rich atmosphereとコンシステント????
%in the envelope of a planet ranges from -0.3 to 0.4, which roughly covers the [Fe/H] range of stars harboring close-in gas giants.
In this study, the metallicity-magnetic field strength relation of close-in gas giants with $Z \lesssim 0.1$ was discussed.
The {\tt MESA} code is not applicable to the thermal evolution of a planet with a metal-rich envelope of $Z \gtrsim 0.1$ because of the incompleteness of the EoS for heavy material, which affects the cooling history of a gas giant \citep{2008A&A...482..315B}.
A large increase in opacities in a metal-rich envelope delays the cooling sequence of a planet. Both the increase in the molecular weight and a reduced adiabatic temperature gradient due to chemical reactions in a metal-rich envelope lower the local pressure, leading to hastening of the envelope contraction \citep[e.g.][]{2011MNRAS.416.1419H,2015A&A...576A.114V}. These effects on magnetic field strengths of close-in planets having metal-rich envelopes will be a topic for future work.

%\begin{figure}
%    \centering
%    \includegraphics[clip,scale=0.5]{fig_Bs-massive.eps}
%    \caption{Same as Figure \ref{fig:Bdy-massive} but for the magnetic field strength on the planetary surface at the pole ($B^{\rm pole}_{\rm s}$).}
%    \label{fig:Bs-massive}
%\end{figure}

%\begin{figure}
%    \centering
%    \includegraphics[clip,scale=0.5]{fig_Bs-low.eps}
%    \caption{Same as Figure \ref{fig:Bdy-low} but for $B^{\rm pole}_{\rm s}$.}
%    \label{fig:Bs-low}
%\end{figure}

\section{Discussion} \label{sec:discussion}

The radio power emitted by a planet is believed to come from the kinetic energy of particles incident on the magnetopause.\footnote{The radio emission can also be generated by the interaction between charged particles from a star and the interplanetary magnetic field \citep{2001Ap&SS.277..293Z}}.
Assuming that the emitted radio power ($P_{\rm rad}$) is proportional to the total kinetic energy flux of energetic charged particles \citep{1984Natur.310..755D}, $P_{\rm rad}$ along the planetary magnetosphere is expressed by $P_{\rm rad} = m n v^3 \pi R_{\rm s}^2$,
where $m$ is the mass of a particle, $n$ is the number density of particles from the stellar wind and coronal mass ejections, $v$ is the particle velocity, and $R_{\rm s}$ is the radius of the planetary magnetosphere.
A promising candidate of a radio emitter is close-in planets that are exposed to a high flux of energetic particles from the star, although plasma in a dense stellar wind may obscure the radio emission from a planet.
Superflares on low-mass stars occur more frequently than those on Sun-like stars \citep[e.g.][]{2014IAUS..293..393M}. Very strong radio emissions from close-in gas giants may be observed sporadically around low-mass stars.
The size of the planetary magnetosphere is defined as the standoff distance where the pressure of a planetary magnetic field is balanced by the sum of the ram pressure of a stellar wind and the thermal pressure of electrons and protons at the substellar point,\footnote{$m n v^2 + 2 n kT \sim \frac{\mu_0 {\cal M}^2}{8\pi^2 R^6_{\rm s}}$, where $T$ is the temperature of a particle, $k$ is the Boltzmann constant, and the magnetic dipole moment of a planet $\cal{M}$ is given by $B^{\rm pole}(r)/2\mu_0 \sim {\cal M}/(4\pi r^3)$, where $B^{\rm pole}(r)$ is two times as large as the magnetic field strength at the equator.} namely, $R_{\rm s} \propto B^{1/3} R_{\rm p}$, where $B$ is the magnetic flux density on the planetary surface $R_{\rm p}$ at the pole.
A young planet close to a star is expected to have a strong radio emission because of a large radius of the planet and a strong dynamo-driven magnetic field (see Figure \ref{fig:Bdy-massive}).
The magnetic field strength of a hot Saturn is strongly correlated to its core mass (see Figure \ref{fig:Bdy-low}). Non-detection or very weak signals of the auroral radio emission from a young, hot Saturn (the mass $\lesssim 0.3\,M_{\mathrm{Jup}}$ and its age $\lesssim 0.1\,\mathrm{Gyr}$) is suggestive of the existence of a hidden small core. The expected radio power emitted by a hot Saturn is, however, lower by one order of magnitude than those emitted by hot Jupiters
unless metallization of hydrogen occurs at $\lesssim 1-1.5$\,Mbar.
Besides, a weak radio emission from a hot Saturn may be shielded by the plasma in the Earth's ionosphere because of an $f_\mathrm{cyc} < 10\,\mathrm{MHz}$. 
Thus, close-in gas giants with $\gtrsim 0.4\,M_\mathrm{Jup}$ are good targets for ground-based radio telescopes to search for radio emissions form exoplanets (see \citet{2016ApJ...820..122F} and \citet{2017pre8.conf..285G} for the detectability of radio emissions from hot Jupiters).

In this paper, the interiors of giant planets were assumed to have layered structures. Recent first-principle calculations of the solubility of core material in metallic hydrogen suggested that both rocky material (MgO) and water ice are highly soluble in metallic hydrogen near the core-envelope boundary of Jupiter and Saturn \citep{2012ApJ...745...54W,2012PhRvL.108k1101W}.
If core erosion occurred deep inside irradiated giant planets after their formation,
the emergence of dynamo-driven magnetic fields could be delayed in hot Saturns with diluted cores.
The heavy-element pollution due to planetesimal or pebble accretion as well as the dredge-up of core material may produce non-negligible compositional gradients in the envelope. 
Since a sharp compositional gradient inhibits large-scale overturning convection, double-diffusive convection may develop in the envelope of a giant planet and reduce the efficiency of the heat transport \citep{2012A&A...540A..20L}. 
If the mixing of heavy elements in the convective envelope of a giant planet, however, occurs efficiently, the initial compositional distribution in the envelope of a giant planet has no strong impact on its long-term thermal evolution \citep{2015ApJ...803...32V}.

The energy-flux-based scaling law of a planetary magnetic field was applied to irradiated giant planets. A highly irradiated gas giant is likely to be tidally locked by its parent star. A tidally locked gas giant should be a slow rotator with a rotation period ($P_\mathrm{rot}$) of about a few to 10 days, compared to Jupiter and Saturn with $P_\mathrm{rot} \sim 10\,\mathrm{hrs}$.
The magnetic field strength of a slowly rotating planet may be weaker than that predicted by equation (\ref{eq:ave-B}). 
The applicability of such a scaling law to a tidally locked gas giant should be validated by future dynamo simulations. 
We here emphasize that the emergence of a dynamo region in a Saturn-mass planet is controlled by the existence of metallic hydrogen.

%The flux density $\Phi$ of the auroral radio emission from a planet that we can observe at the Earth is given by $\Phi = P_{\rm rad}/(\Omega d^2 \Delta f)$, where $\Omega$ is the solid angle of the emitted radio beam, $d$ is the distance between a planet and the Earth, and $\Delta f$ denotes the bandwidth of the radio emission. Then we obtain the relation of $\Phi \propto B^{-1/3} R^2_{\rm p} d^{-2}$, using $\Delta f = f^{\rm max}_{\rm cyc}$ \citep{2007P&SS...55..618G}, where $f^{\rm max}_{\rm cyc}$ is the maximum cyclotron frequency. 

\section{Summary}
Close-in gas giants are expected to have a strong magnetic field of $\sim 10-100$\,G. Future radio observations in the frequency range of $\gtrsim 10$\,MHz and the spectropolarimetry of atomic emissions such as the triplet He I are capable of detecting the magnetic field of exoplanets. In this paper, we have examined the linkage between the existence of innermost cores of extrasolar giant planets and planetary magnetic fields. We have simulated the long-term thermal evolution of a $0.2-10\,M_\mathrm{Jup}$ irradiated gas giant with no core or a core of $1-10\,M_\oplus$ using the {\tt MESA} code. A young, massive close-in gas giant tends to have a strong magnetic field. The magnetic field strength of a hot Jupiter is insensitive to its core mass, whereas the core mass strongly affects the emergence of a planetary dynamo in a hot Saturn. 
If metallization of hydrogen occurs at $> 1-1.5$\,Mbar,
no dynamo-driven magnetic field is generated in a Saturn-mass planet with no core or a small core until $\sim 10-100$\,Myr.
A highly luminous, massive gas giant that receives a lower incident flux from its parent star produces a stronger magnetic field because it has to radiate away a large amount of internal heat to achieve the thermal equilibrium. The magnetic field strength of a thermally evolved gas giant after $\sim 100\,\mathrm{Myr}$ is almost independent of the stellar incident flux. Although the magnetic field strength is a good indicator of the core inside a young, hot Saturn with $\lesssim 0.3M_\mathrm{Jup}$, constraining on the interior structure through the radio detection may be challenging because of the weakness of radio signals and the shielding effect of plasma in the Earth's ionosphere. Hot Jupiters with $\gtrsim 0.4\,M_\mathrm{Jup}$ can be promising candidates for future ground-based radio observations such as the SKA-low.

\acknowledgments

YH is supported by a Grant-in-Aid for Scientific Research on Innovative Areas (JSPS/KAKENHI grant No. 18H05439). We thank the anonymous referee for useful comments and improving our paper.






\appendix

\section{Estimating planetary magnetic field strengths from {\tt MESA} calculations}

The energy-based scaling rule for a dynamo-driven magnetic field in a gas giant (see equation (\ref{eq:ave-B})) is related to four thermodynamic properties: the convective heat flux $q_\mathrm{c}$, the density $\rho$ and the characteristic length scale $L$ in the convective layer, and the temperature scale height $H_T$,
where $L$ is given by $\min(D, H_\rho)$, $D$ is the thickness of the convective layer, and $H_\rho$ is the density scale height.
Two scale heights are given by $H_\rho = P(\partial \ln{P}/\partial \ln{\rho})_S/(\rho g)$ and $H_T = P/(\rho g\nabla_{\rm ad})$, where $g$ is the gravitational acceleration, $P$ is the pressure of a fluid, $S$ is the specific entropy, and $\nabla_{\rm ad}$ is the adiabatic temperature gradient.
%Two scale heights are given by $H_\rho = - dr/d\ln{\rho} = P(\partial \ln{P}/\partial \ln{\rho})_S/(\rho g)$ and $H_T = c_p/(\alpha g) = P/(\nabla_{\rm ad} \rho g)$, $\alpha$ is the thermal expansivity, and $g$ is the gravitational acceleration, respectively.
{\tt MESA} calculations determine the internal profile of a planet (e.g., $P(r)$, $T(r)$, and $\rho(r)$) at a given time and also give $c_p$, $\nabla_\mathrm{ad}$, and derivatives such as $(\partial \ln{P}/\partial \ln{\rho})_S$, $(\partial \ln{P}/\partial \ln{\rho})_T$, and $(\partial \ln{P}/\partial \ln{T})_\rho$ required for $H_T$ and $H_\rho$, where $c_p$ is the specific heat capacity, $T$ is the temperature of a fluid, and $r$ is the radial distance from the center of a planet.

The convective region in the metallic hydrogen envelope of a gas giant, namely, the dynamo region, is determined by the Schwarzschild criterion for convective instability, which gives the thickness of a convective layer $D$.
The convective heat flux, $q_\mathrm{c}(r)$, is determined by $v_\mathrm{conv}$ given in 
the default outputs of {\tt MESA}.
Following the mixing length theory \citep[e.g. see][]{1990sse..book.....K}, we obtain the average velocity of convective elements $v_{\rm conv}$:
\begin{equation}
    v_{\rm conv}^2 = g \delta (\nabla - \nabla_{\rm e}) \frac{{\ell}^2}{8H_P},
\end{equation}
where $\delta = - (\partial \ln{\rho}/\partial \ln{T})_P$, 
$\nabla$ is the actual temperature gradient, $\nabla_{\rm e}$ is the temperature gradient of fluid elements, and $H_P$ is the pressure scale height given by $P/(\rho g)$. We use the mixing length $\ell = 2H_P$. Then the convective heat flux $q_\mathrm{c}$ is expressed by
\begin{equation}
     q_\mathrm{c} = \frac{\rho c_P T}{4\sqrt{2}} \sqrt{g \ell \delta}\left(\frac{\ell}{H_P}\right)^{3/2}
        (\nabla - \nabla_{\rm e})^{3/2} = \frac{2 c_P T \rho^2 v^3_{\rm conv}}{P \delta}.
    \label{eq:q}
\end{equation}
Substituting $v_\mathrm{conv}$ and other thermodynamic properties given by {\tt MESA} into equation (\ref{eq:q}), we obtain $q_\mathrm{c}(r)$ and $q_0$ in equation (\ref{eq:ave-B}). The time-varying $F$ in equation (\ref{eq:ave-B}) can be integrated over the dynamo region at every time step using $L(r)$, $\rho(r)$, $q_0$, $q_\mathrm{c}(r)$, and $H_T(r)$. Thus, the magnetic field strength of a close-in gas giant is estimated from {\tt MESA} calculations based on the energy-based scaling rule in equation (\ref{eq:ave-B}).

\bibliography{reference}{}
\bibliographystyle{aasjournal}



\end{document}